\documentclass[conference]{IEEEtran}
\ifCLASSINFOpdf
  \usepackage[pdftex]{graphicx}
  \graphicspath{{./img/}}
\else
  \usepackage[dvips]{graphicx}
  \DeclareGraphicsExtensions{.eps}
\fi

\begin{document}
%
\title{Taking back control of HPC file systems with Robinhood Policy Engine}

\author{\IEEEauthorblockN{Thomas Leibovici}
\IEEEauthorblockA{CEA, DAM, DIF, F-91297 Arpajon, France \\
Email: thomas.leibovici@cea.fr}}

\maketitle

\begin{abstract}
Today, largest Lustre file systems store billions of entries.
On such systems, classic tools based on namespace scanning become unusable.
Operations such as managing file lifetime, scheduling data copies and generating overall
filesystem statistics become painful as they require collecting, sorting and
aggregating information for billions of records. \\
\emph{Robinhood Policy Engine} is an open source software developed to address these challenges.
It makes it possible to schedule automatic actions on huge numbers of filesystem entries.
It also gives a synthetic understanding of file systems contents by providing overall statistics about data ownership, age and size profiles.
Even if it can be used with any POSIX filesystem, Robinhood supports Lustre specific features like OSTs, pools, HSM, ChangeLogs, DNE...
It implements specific supports for these features, and takes advantage of them to manage Lustre file systems efficiently.
\end{abstract}

\section{Introduction}

The largest filesystems in HPC now reach tens of petabytes\cite{Seq12}
and exabyte-sized storage systems will emerge by the end of the
decade\cite{Exa}. Operating such filesystems is not only a challenge for data
management, it is also a matter of handling metadata.

As systems become bigger, compute codes continue to organize their
information in a traditional manner, by crea\-ting files and directories in the
filesystem namespace. As a consequence, growing filesystem capacities
inexorably result in larger namespaces and in dramatically increasing number
of metadata objects (hundreds of millions to billions).

Conventional tools that massively query filesystem metadata (like {\tt find},
{\tt du}, {\tt rsync}...) mostly use POSIX\cite{POSIX} name\-space scanning, which
consumes a lot of time and operates very slowly on large namespaces.

\emph{Robinhood Policy Engine}, an OpenSource project, has been developed to
address these issues. It aims to efficiently collect information using parallel
scanning me\-cha\-nisms, or, when such a feature is available, monitor incremental changes from a filesystem,
thus alleviating the need for full namespace scans.

The collected information is stored into a database, which mirrors filesystem
metadata. This auxiliary database offers many possibilities:

\begin{itemize}
\item Searching for entries using various criteria, much more efficiently than
using the POSIX API. For instance, the following request can be handled more efficiently by a database
than a filesystem:\\
{\tt \small select * from ENTRIES where size < 1024}\\
vs.
{\tt \small find /fs -size -1024}
\item Extracting accurate and customizable statistics about the
filesystem contents.
\item Massively applying policies on filesystem entries,
based on various criteria like file attributes, path, extended attributes...
\end{itemize}
All these metadata queries do not generate extra load on the
filesystem as they are performed directly on the database.
\\ Moreover, when Robinhood reads incremental changes
from a filesystem in soft real-time,
the query result is immediate and up-to-date, unlike the result of a traditional
scan-based tool that would complete only after hours or days and reflect a past
state of the filesystem.

\emph{Robinhood Policy Engine} has a growing popularity in the Lustre community,
especially since Lustre 2.x releases, as it supports new valuable Lustre features
like MDT Changelogs and HSM. \\
Besides this important users community, Robinhood is now integrated by several vendors
to their Lustre software distributions (like Bull, Cray, Intel...). These vendors are now major constributors to the project.


\section{Robinhood in a nutshell}

\subsection{Big picture}

\begin{figure}[here]
\centering
\includegraphics[width=3.5in]{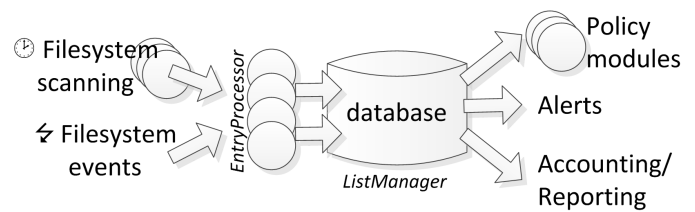}
\caption{\it Architecture overview} \label{fig:BigPicture}
\end{figure}

The concept of \emph{Robinhood Policy Engine} is quite simple: on the one hand, it collects
information from the filesystem it monitors and inserts this information into a
database; on the other hand, it uses the database contents to schedule actions
and provides various metrics and consolidated views of the filesystem.\\
Figure~\ref{fig:BigPicture} gives an overall view of its components.

\subsection{Main features}
\subsubsection{Policies}
\emph{Robinhood} implements various policies to archive data, release disk space,
or remove directories. Po\-li\-cy rules can be specified using multiple conditions on file attributes,
like path, owner, size, last access time...

Here is an example of expression to match a given set of entries:
{\tt \small
\begin{verbatim}
    (size > 1GB or owner == 'foo')
    and path == /my/fs/*.tar
\end{verbatim}
}

\subsubsection{Alerts}
Monitoring filesystems usage is a key point to preserve
the quality of service of a system. \emph{Robinhood} makes it possible to define
alerts on filesystem entries to detect abnormal or toxic behaviors. When
detecting such an entry, it triggers a configurable action such as sending an
email or logging to an alert file.

\subsubsection{Statistics}

\emph{Robinhood} provides detailed statistics about filesystems and gives an
accurate representation of its contents' characteristics.

Commonly used statistics are pre-generated in the database. They are computed
on-the-fly as entries are updated, so the following information is always
available: statistics per object type, per user, per group,
per migration status (for archiving systems) and file size profile. Ranking
"top" users by inode count, by volume, by average file size, by percentage of
files in a given size range is also immediate.\\
For example, getting the following information is a O(1) operation on
the database:
{\tt \small
\begin{verbatim}
# rbh-report -u foo
user,     type,   count,   spc_used,  avg_size
foo ,      dir,     261,    1.02 MB,   4.00 KB
foo ,     file,   17121,   20.20 TB,   1.21 GB
foo ,  symlink,       4,   12.00 KB,        61
\end{verbatim}
}

\begin{figure}[here]
\centering
\includegraphics[width=45mm]{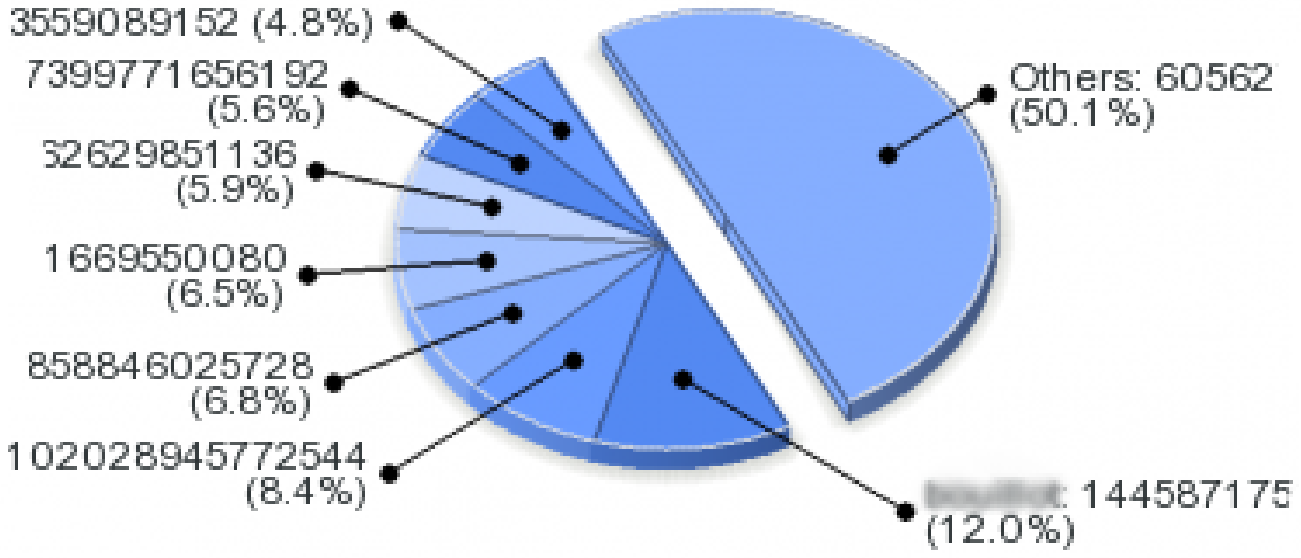}
\includegraphics[width=42mm]{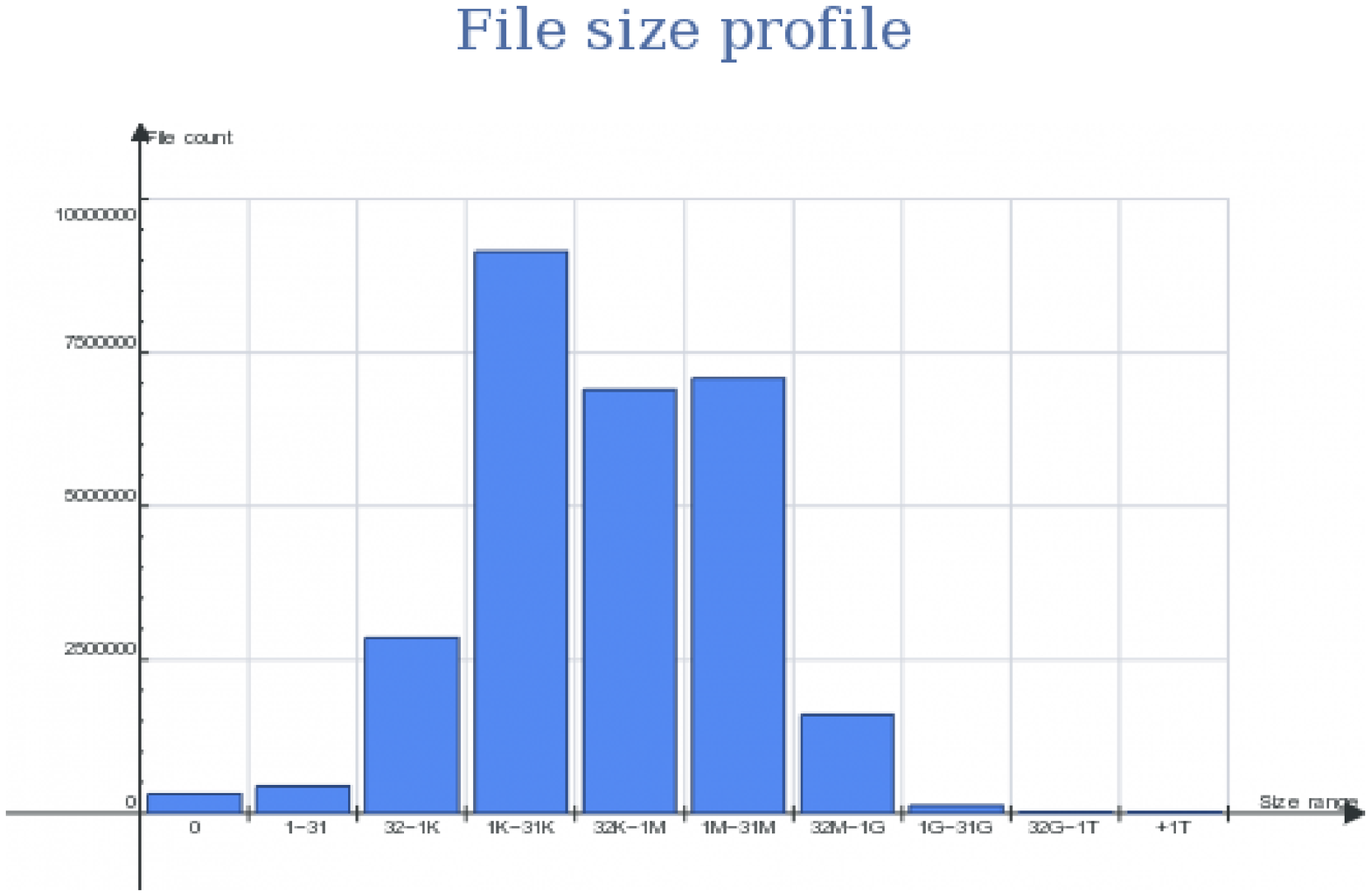}
\caption{\it Web interface overview: space usage and file size profile}
\label{fig:szprof}
\end{figure}

All those statistics are also available in the web interface.
Figure~\ref{fig:szprof} shows the repartition of space usage and files sizes
profile for a given user in the web interface.

In addition to summary reports, lists of particular entries can also be queried:
top directories sorted by any criteria like inode count or average file size,
top largest files, oldest files, and so on.

\subsubsection{{\tt du} and {\tt find} clones}

\emph{Robinhood} provides enhanced clones of traditional UNIX {\tt find} and {\tt du} commands.
These commands query the robinhood database instead of scanning the filesystem, which makes them faster.

\subsection{Lustre specific features}
\emph{Robinhood} can run on any POSIX filesystem.
However, it implements specific features for Lustre filesystems.
\subsubsection{OSTs and pools}
Lustre parallelizes access to data
across multiple vo\-lumes called {\it OST}s (Object Storage Targets).
\emph{Robinhood} is able to independently monitor the usage of these OSTs and balance
disk usage between them: if one of them exceeds a given threshold,
\emph{Robinhood} can apply purge policies targeted to the files located on that
particular OST.

In the same fashion, it can control the usage of Lustre OST pool, which are
administratively-defined groups of OSTs.

OST index and pool name can also be used as a criteria in policy definitions.

\subsubsection{MDT ChangeLog}
\emph{MDT ChangeLog} is an available feature since
Lustre 2.0. It consists in logging metadata change operations to a transactional
and persistent log. A user-space process can register as a log consumer,
to be aware of the changes in the filesystem (file creation, rename, unlink,
chmod, ...). Changelog records are kept on persistent storage until the
consumer reads and acknowledges them. Thus, no event can be lost, even if the
consumer is not running.

\emph{Robinhood} can read Lustre MDT ChangeLog. When processing a change record,
it acknowledges it only after the related change has been committed to its own
database. Thus, the transactional and persistent aspects of event processing are
preserved. Using this mechanism, \emph{Robinhood} maintains a replicate of
filesystem metadata which is updated in soft real-time by reading the ChangeLog.
Scanning the filesystem is not required anymore in order to update the database.

\subsubsection{Lustre-HSM}

\emph{Lustre-HSM} feature allows using a Lustre filesystem as the top level of a storage hierarchy -- in front of a HSM\footnote{Hierarchical Storage Manager}, to benefit from both Lustre high performances and HSM large and cheaper storage resources.
Implementing such a mechanism requires to monitor filesystem contents
and disk space usage, to archive data and to make room in the filesystem when it fills up.

As it implements all the needed features, \emph{Robinhood} can be used as a \emph{Policy Engine} for this cache. In this mode, it monitors HSM specific events from MDT Changelog, release unused files data when space is lacking on OSTs, and trigger file archiving requests. Data retrieval is handled automatically by Lustre.

Using \emph{Robinhood} policies, data is moving to HSM and back to Lustre
depending on file access patterns, so that only {\it hot} data is kept on Lustre.
\emph{Lustre-HSM} also benefits from the {\it undelete} and {\it disaster recovery} features of \emph{Robinhood}.

\section{Challenges}
Robinhood has to address multiple challenges to scale on largest filesystems: scan fast,
process a high throughput of information, store and query this information efficiently.
And as filesystems become more performant and implement more parallelism,
Robinhood must implement new solutions to stay in the race.

\subsection{Collecting information}
\subsubsection{Scanning}

Even if using the Lustre Changelog me\-cha\-nism, an initial scan is still needed to populate robinhood database with the initial filesystem state. This section describes the implemented solutions to make this scan as faster as possible, and future directions that could be considered to make it even faster.

\begin{figure}[here]
\centering
\includegraphics[width=40mm]{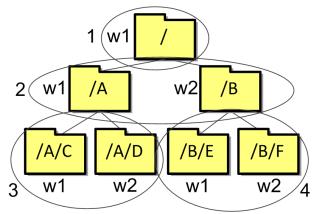}
\caption{\it Parallel traversal with depth-first priority \newline{(example with 2 worker threads: w1 and w2)}}
\label{fig:dft_multi}
\end{figure}

To go beyond the performance of classical scanning tools, robinhood implements a multi-threaded version of depth-first traversal\cite{Knuth}. To parallelize the scan, the namespace traversal is split into individual tasks that consist in
reading single directories. A pool of worker threads performs these tasks following a depth-first strategy (as illustrated on figure~\ref{fig:dft_multi}).

Even if multi-threading improves the scan performance, it is still limited by Lustre single client performance.
To overtake it, robinhood also allows splitting the namespace scan across multiple clients, thus cumulating their RPC throughputs.  In this case, each client runs a robinhood instance. Each instance scans a distinct part of the namespace using the parallel algorithm described above, and they all feed a common database.

However, these implementations still suffer from POSIX. Other solutions could be considered to get rid of it:
\begin{itemize}
\item The POSIX scan could be replaced by a low-level scan of the MDS based on {\tt e2scan}. Unfortunatly, this does not provide all the information about filesystem entries, like file size which is distributed across OSTs. Moreover, such an implementation would be very dependant of MDS storage format and other Lustre internals.
\item Another possible solution would be to implement such a low-level scan as a filesystem service, similar to the changelogs.
The consumer would open a spe\-cial changelog stream that consists of the list of all filesystem entries. The format of this stream would be standard and persistant across Lustre version, and would not depend on Lustre internals.
\end{itemize}

\subsubsection{Processing incoming information}
A key point in Ro\-bin\-hood performance is its ability to process
incoming in\-for\-ma\-tion at a high rate, like entries from a filesystem
scan, or filesystem change events.  The processing requires to access different
resource types (database, filesystem, changelog stream...)
that have different concurrency constaints.

The implemented mechanism consists in splitting record processing into multiple steps, one step for each kind of operation (database, filesystem...).  These tasks are performed in parallel by a pool of worker threads, which allows a fast processing. The load and the concurrency level on the database and the filesystem can be controlled by limiting the number of simultaneous operations of each type processed by the workers.

This mechanism may be improved in the future by making it asynchronous:
the changelog processing would just ``tag'' entries in the database with a set of ``dirty'' attibutes that need to be refreshed.
Then, a pool of ``updaters'' would refresh attributes of the tagged entries in background.
This way, less operations needs to be performed synchronously when processing the changelog records, thus resulting in higher processing rates. Moreover, if many changes occur on a given filesystem entry, it could be tagged multiple times before its attributes are effectively updated, thus reducing filesystem calls and attributes updates in the database.

\subsection{Storing information}

The choice as been made to store information in a tran\-sac\-tional MySQL database to benefit from all its features:
in\-for\-ma\-tion persistency, memory cache management, flexible querying using the SQL language, transaction and concurrency management, backups...
Moreover, a database is more adapted to multi-criteria querying and information aggregation than a filesystem.
From the performance point of view, such an engine can handle hundreds of thousands requests per seconds, which is enough in most cases to handle operations from a Lustre MDS.

However, with the implementation of a distributed name\-space in Lustre (DNE), this single host database model reaches a limit.
As a single database server must handle the workload from multiple MDS, it can become a bottleneck.
To face this challenge, a future direction is to distribute robinhood database. This could be done at software level by splitting incoming information to multiple databases. Another solution is to use a database engine that natively implement such a {\it sharding} feature,  like MongoDB\cite{Mongo}.

\subsection{Reporting aggregated information}

It is useful to aggregate the huge amount of information stored in robinhood database,
to provide meaningful information to system administrators about filesystem contents,
like age profile, size profile, user accounting...

Aggregating and organizing millions or billions of records can be very expensive (several minutes to hours),
but administrators sometimes need to get the information instantly: to track filesystem activity in real-time, to control user usage before submitting a job...
To achieve this, robinhood maintains some pre-aggregated information updated on-the-fly. This information is updated when robinhood processes incoming records, so it is immediatly available when the administrator needs it.
For instance, the following information can be retrieved instantly from robinhood DB and is updated in real-time:
total volume and entry count for each user, group, object type (file, directory...), HSM status,
file size profiles, changelog counters for each type of operations...

In a near future, new statistics could be added to meet the requirements of filesystem administrators:
\begin{itemize}
\item Usage counters for a given level of sub-directories, so commands like {\tt du} will be made instantaneous at this level of the namespace.
\item Per user changelog counters, to track individual user activity.
\item Per jobid changelog counters\footnote{Since Lustre 2.7, the 'jobid' is integrated to changelog records.}, to track jobs activity.
\end{itemize}

Maintaining these counters on-the-fly has a cost: this significantly impacts the changelog processing rate.
A possible solution to avoid this would be to update those counters asynchronously, in background.
As a consequence, returned statistics could be a little outdated compared to the effective filesystem content, but they would still be updated near real-time, which is acceptable for most use cases.

\subsection{New file system architectures}

In the past, most of Lustre filesystems were homogeneous and all OSTs consisted of spinning disks.
A trend in filesystems architecture is now to combine technologies like SSDs and spinning disks,
to benefit from both SSD throughput and disk capacity with a limited cost.

Such architectures create new needs in data management. Data must be moved between pools
of storage resources according to site-specific policies, like in a HSM.
Moreover, these data movements between pools can be used together with the existing Lustre HSM feature,
which requires a coordination between data management policies.

\begin{figure}[here]
\centering
\includegraphics[width=3.5in]{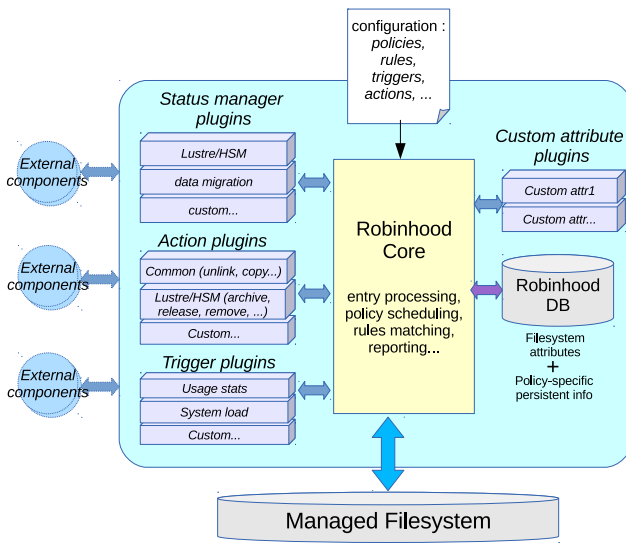}
\caption{\it Robinhood v3 plugin-based architecture}
\label{fig:rbhv3}
\end{figure}

To satisfy these requirements, robinhood must be adapted to manage more and more policies in a single instance,
and to allow coordination between the policies. This is the goal of a major ongoing development in robinhood: {\it generic policies}.
Thanks to this new feature, administrators will be able to schedule any kind of action on filesystem entries, including (but not restricted to) all "legacy" policies, internal data migration in Lustre, data integrity checks, post-processing... Administrators can use plugins shipped with robinhood to define custom policies by simply writing a few lines of configuration. They can also develop their own plugins to implement specific mechanisms. \\
This major robinhood evolution will be part of the upcoming robinhood v3 (big picture represented in fig.~\ref{fig:rbhv3}).

\vfill
\break

\section{Conclusion}

\emph{Robinhood Policy Engine} is a complete, integrated and efficient solution
for performing common administrative tasks on large filesystems, to closely
monitor their contents and to integrate them into hierarchical storage
architectures.
It is continuously adapted to support and take advantage of new Lustre features, and to satisfy administrators needs. In par\-ti\-cu\-lar, major evolutions are in progress to break scaling barriers and address new requirements in terms of performance and data management.


%
\bibliographystyle{IEEEtran}
\bibliography{IEEEabrv,./bibliography}
%
%

\end{document}